\documentclass[14pt,twoside,a4paper]{article}
\usepackage[noadjust]{cite}
\usepackage{indentfirst}
\usepackage{amsmath}
\usepackage{amsfonts}
\usepackage{amssymb}
\usepackage{graphicx}
\usepackage{subfig}
\usepackage{authblk}

\title{\textbf{Dynamical theory of diffraction on moving grating}}
\author[1]{V.A. Bushuev \thanks{e-mail: vabushuev@yandex.ru}}
\author[2]{A.I. Frank}
\author[2]{G.V. Kulin}
\affil[1]{\textit{Lomonosov Moscow State University, Moscow, Russia}}
\affil[2]{\textit{Frank Laboratory of Neutron Physics, Joint Institute for Nuclear Research, Dubna, Russia}}
\date{}

\begin{document}
\maketitle
\bibliographystyle{unsrt} 

\begin{abstract}
In the framework of the approximation of slowly varying amplitudes a multiwave dynamical theory of neutron diffraction on a moving phase grating was developed. The influence of the velocity of the grating, its period and height of the slits on the discrete energy spectrum and intensity of various diffraction orders was analyzed.
\end{abstract}

\section{Introduction}
In \cite{PhysLettA1994} it was predicted that a diffraction grating moving across the beam of slow neutrons, is a time-dependent quantum modulator that converts the energy spectrum of incident neutrons in a series of quantum states with discrete energies. Later a discrete spectrum in the diffraction of ultracold neutrons (UCN) on a moving grating was observed in the experiments \cite{PhysLettA2003,JinrCommun2004,JETPLett2005,JETPLett2007}. In addition, it was shown that a moving grating with a varying spatial period can serve as a quantum time lens for UCN \cite{JETPLett2003}.

\begin{figure}[!h]
\centering
    \subfloat[\label{fig:rest}]{\scalebox{0.6}{ \includegraphics[page=1]{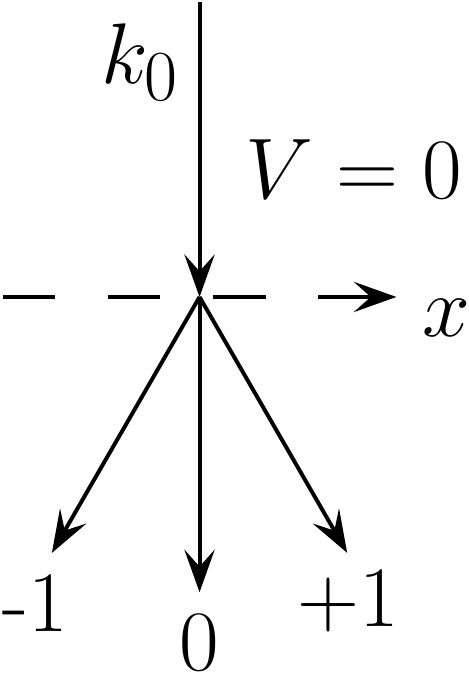}}}
    \subfloat[\label{fig:movingframe}]{\scalebox{0.63}{ \includegraphics[page=2]{pdfpics.pdf}}}
    \subfloat[\label{fig:labframe}]{\scalebox{0.52}{ \includegraphics[page=3]{pdfpics.pdf}}}
    \caption{Schemes of diffraction. a – grating at rest,$\omega_{j} = \omega_{0}$; b – coordinate system moves together with the grating $\omega_{j}=\omega^\prime$; c – moving grating in the laboratory coordinate system $\omega_{j} = \omega_{0}+j\Omega$.
}
\label{fig:diffractions}
\end{figure}

Let us briefly present the result of the corresponding quantum problem. In the laboratory coordinate system a plane neutron wave $\Psi_{in}(x,z,t) = A_{0} exp(ik_{0}z-i\omega_{0}t)$ is assumed to propagate along the z-axis towards a periodic grating, which is supposed to be normal to the neutron propagation axis (Fig.~\ref{fig:rest}), where $k_{0} = mV_{0}/\hbar$ is the neutron wave number, $m$ is the neutron mass, $\hbar$ is the Planck constant, $\omega_{0} = \hbar k_{0}^{2}/2 m$. The grating moves with a constant velocity $V$ in the positive direction along the $x$ axis (Fig.~\ref{fig:movingframe}) and the grating grooves are oriented along the $y$ axis. As in \cite{PhysLettA1994,PhysLettA2003,JinrCommun2004,JETPLett2005} let us solve the problem in the moving coordinate system $(x^\prime, z)$ where the grating is at rest. In this system the wave is obliquely incident on the grating:
\begin{equation}
\Psi^\prime_{in}(x^\prime,z,t) = A_{in} exp(ik_{0}z-i\omega^\prime t),
\label{eq:psimovframe}
\end{equation}
with $A_{in}(x)=A_{0}exp(ik_{V}x)$, $k_{V} = mV/\hbar$,  $\omega^\prime= \omega_{0}+\omega_{V}$, $\omega_{V} = \hbar k_{V}^{2}/2 m$. After the passage of the grating in the region $z>$0 we have
\begin{equation}
\Psi^\prime(x^\prime,z,t) = A_{0}\int\limits_{-\infty}^{\infty} F(q) exp(iqx^\prime +ik_{z}z-i\omega^\prime t)\,\mathrm{d}q,
\label{eq:psimovframeint}
\end{equation}
where $k_{z} = (k_{0}^{2} + k_{V}^{2}-q^{2})^{1/2}$, $F(q)$ is the Fourier transform of the product $F(x) = A_{in}(x)f(x)$, where $f(x)$ is the grating transmission function. For a periodical function $f(x) = \sum_{j}a_{j}exp(iq_{j}x)$ where $j$ are integers, $q_{j} = 2 \pi j/d$, $d$ is the spatial period of the grating, $F(q)=A_{0}\sum_{j}\delta(q-q_{j}+k_{V})$, and the Fourier coefficients are
\begin{equation}
a_{j} = \frac{1}{d}\int\limits_{0}^{d} f(x) exp(iq_{j}x)\,\mathrm{d}x,\; j=0,\pm 1,\pm 2,\ldots
\label{eq:ampl}
\end{equation}
Returning to the laboratory coordinate system $x = x^\prime + Vt$ we obtain a superposition of plane waves with amplitudes $a_{j}$, discrete frequencies $\omega_{j}$  and wave vectors $\textbf{k}_{j} = (q_{j}, k_{zj})$ (see Fig.~\ref{fig:labframe}):
\begin{equation}
\Psi(x,z,t) =\sum_{j} a_{j} exp(iq_{j}x+ik_{zj}z-i\omega_{j} t),
\label{eq:psilab}
\end{equation}
where $k_{zj} =(k_{0}^{2}+ 2k_{V}q_{j}-q_{j}^{2})^{1/2}$, $\omega_{j}=\omega_{0} + j\Omega$,
$\Omega=2\pi /T$, $T = d/V$. We should emphasize that the waves with different $j$ do not affect each other.

\begin{figure}[!h]
\centering
    \scalebox{0.7}{\includegraphics[page=4]{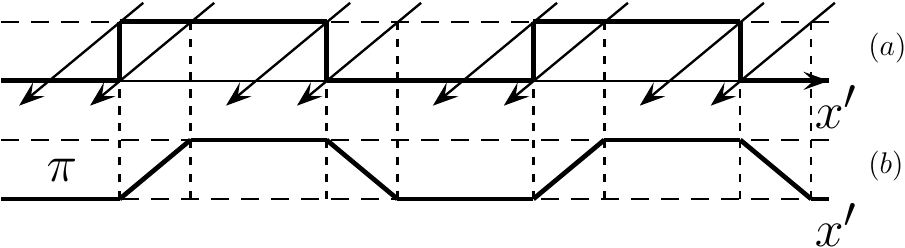}}
    \caption{\textit{a} – grating profile, \textit{b} – phase profile.}
\label{fig:grasinc}
\end{figure}

Let us consider a grating with a rectangular groove profile with width $d/2$ and depth $h$, so that the phase difference $\phi= k_{0}(1-n)h =\pi$, where $n$ is the neutron refraction index. At normal incidence the amplitudes of even orders (including the zero order) are $a_{j}=0$, but odd-order ones are $a_{j} = 2i/\pi j \;(j = 2s-1)$. It is obvious that for larger values of spectral splitting $\Omega= 2\pi V/d$ it is necessary to increase the grating velocity $V$ and (or) decrease the spatial period $d$. But due to the oblique incidence in the moving coordinate system the phase profile varies with increasing velocity and takes the form of a trapezium \cite{JinrCommun2004} (Fig.~\ref{fig:grasinc}). It follows from~\eqref{eq:ampl} that amplitudes of even orders are $a_{j} = cB_{j}$ and those of odd orders are $a_{j} = -B_{j}/j$, where
\begin{equation}
B_{j}=[1+exp(-i\pi jc )] /[i\pi(1 -j^{2}c^{2})],
\label{eq:edgeampl}
\end{equation}
where dimensionless parameter $c = 2hV/dV_{0}$ is introduced \cite{JinrCommun2004}. At $c = 1$ the phase profile takes the form of a triangle. In the case of a moving grating $(c>0)$  the amplitudes of zero and all even orders never vanish and their intensities $\vert a_{j}\vert^{2}$ increase with increasing $c$, contrary to the intensities of odd orders, which decrease (see Fig. 3). In the experiment [5] a decrease in the intensity of minus $1^{st}$ order with increasing velocity of the grating was really observed, but it was greater than that predicted by the relation~\eqref{eq:edgeampl}.

Quantum modulation of neutron waves by a moving grating is a methodological basis of experiments to test the equivalence principle for the neutron \cite{JETPLett2007, NIMA2009}. Therefore a more rigorous theoretical study of the phenomenon is required. In this paper we developed a dynamical theory of neutron diffraction on a moving grating which to some extent satisfies this requirement. It is based on the method of slowly varying amplitudes and takes into account the mutual influence of waves of different orders as the neutron penetrates a 3D phase grating.
\section{Theory}
Neutron propagation in a medium is described by the Schrödinger equation
\begin{equation}
\Delta\Psi(\textbf{r},t)+[k^{2}-k_{b}^{2}(\textbf{r})]\Psi(\textbf{r},t)=0,
\label{eq:shredinger}
\end{equation}
where $\Delta$ is the Laplace operator, $k$ is the wave number of neutrons in vacuum, $k_{b}^{2}(\textbf{r}) = 4\pi N(\textbf{r})b(\textbf{r})$, $N (\textbf{r})$ is the density of nuclei, $b(\textbf{r})$ is the coherent scattering length of the medium. Let us assume that function $k_{b}^{2}(x)$ periodic in region $0\leq z\leq h$ has the form
\begin{equation}
k_{b}^{2}(x)=\sum_{n=-\infty}^{\infty} \chi_{n}exp(iq_{n}x),
\label{eq:kb2}
\end{equation}
where $q_{n} = 2\pi n/d$ are reciprocal lattice vectors,
\begin{equation}
\chi_{n}=\frac{1}{d}\int\limits_{0}^{d} k_{b}^{2}(x) exp(-iq_{n}x)\,\mathrm{d}x.
\end{equation}

A zero Fourier amplitude $\chi_{0}$ determines the value of the average grating refractive index  $n_{e} = (1 -\chi_{0} / k^{2})^{1/2}$ in a layer of thickness $h$.

The wave function of neutrons in the region $0\leq z\leq h$ in the moving coordinate system we can write using~\eqref{eq:psimovframe}, as the sum of the Bloch functions with amplitudes $\Psi_{m} (z)$ depending on the vertical coordinate $z$:
\begin{equation}
\Psi^\prime (x^\prime ,z,t) =\sum_{m=-\infty}^{\infty} \Psi_{m} (z) exp(i \textbf{q}_{m} \textbf{r}^\prime -i\omega^\prime t),
\label{eq:psiprime}
\end{equation}
where the projections of the wave vectors $q_{mx} = q_{m}-k_{V}$, $q_{mz} = k_{0z}= (k_{0}^{2}-\chi_{0})^{1/2}$. Here, it was assumed that the wave number in vacuum is $k^\prime = (k_{0}^{2} + k_{V}^{2})^{1/2}$, and in the layer $0 \leq z\leq h$ it is $k_{0z} = (k^{\prime 2}n_{e}^{2}- k_{V}^{2})^{1/2}$.

By inserting~\eqref{eq:kb2} and~\eqref{eq:psiprime} into~\eqref{eq:shredinger}, we equate the terms with equal exponents and neglect the second derivative of $\Psi_{m}$ with respect to $z$. As a result, we obtain a system of differential equations:
\begin{equation}
\frac{\mathrm{d}\Psi_{m}}{\mathrm{d}z}=i\alpha_{m}\Psi_{m}-i\sum_{n\neq 0}\beta_{n}\Psi_{m-n},
\label{eq:diffeqs}
\end{equation}
where $\alpha_{m} = q_{m}(2k_{V}-q_{m})/2k_{0z}$, $\beta_{n} = \chi_{n}/2k_{0z}$. System of equations~\eqref{eq:diffeqs} should be supplemented by boundary conditions : $\Psi_{0} (z = 0) = A0$, $\Psi_{m\neq 0} (z = 0) = 0$. Neglecting second derivatives when deriving~\eqref{eq:diffeqs}, we excluded the reflected waves from consideration. Apparently, this should not result in serious errors, since for typical experimental conditions \cite{PhysLettA2003,JinrCommun2004,JETPLett2005,JETPLett2007,JETPLett2003} the Fresnel reflection coefficient is rather small: $\vert R_{F}\vert ^{2}\approx 2-4\%$.

It can seen from~\eqref{eq:diffeqs} that unlike the simple model described above, the amplitudes of waves of arbitrary $m$-th order depend on the amplitudes of other orders. Coupling coefficients $\beta_{n}$ in addition to the order number are also defined by material and shape of the grating profile. In our case $\chi_{n} = k_{b}^{2}sin (\pi n / 2 ) / \pi n$. The $\alpha_{m}$ values depend on the grating velocity and the diffraction order. They vanish both at m = 0 and Bragg condition $2k_{V} = q_{m}$. The system of equations~\eqref{eq:diffeqs} is solved using the Runge-Kutta $4^{th}$-order method.

In the laboratory coordinate system, the wave function is also defined by~\eqref{eq:psilab} in which it is necessary to replace amplitudes $a_{m}$ with $\Psi_{m} (z = h)$ from~\eqref{eq:diffeqs}. Account should be taken of the fulfillment of the condition $q_{m} (q_{m}-2k_{V})\leq k_{0}^{2}$ , which excludes exponentially attenuated (evanescent) waves and imposes certain restrictions on the admissible values of diffraction orders $m$ and grating velocity $V$. In practice, diffraction of neutrons on a moving grating takes place when they pass through a rotating disk \cite{PhysLettA2003,JinrCommun2004,JETPLett2005,JETPLett2007,JETPLett2003,NIMA2009} that has radial rectangular grooves with a period $d$ and depth $h$. If $R$ is the distance of a ring with grating structure from the center of the disk and $f_{d}$ is its rotation frequency, then $V = 2\pi Rf_{d}$.

If the Bragg condition $\alpha_{1} = 0$ is fulfilled and all orders except $m = 0$ and $m = 1$  can be ignored, then~\eqref{eq:diffeqs} is reduced to the well-known X-ray crystal optics Takagi equations in the Laue geometry, whose solution has the following simple form:
$\Psi_{0} (z) = A0 cos(\pi z / L_{ex}),\; \Psi_{1} (z) = -i A0 sin(\pi z / L_{ex})$.
Here $L_{ex} = \pi 2k_{0z} / k_{b}^{2}$ is the so-called extinction length at which the radiation from the zero-order fully pumped into the first order and vice versa. Under conditions of the experiments \cite{PhysLettA2003,JinrCommun2004,JETPLett2005,JETPLett2007,JETPLett2003,NIMA2009}, it is a fraction of a micrometer which is comparable with the depth of the grating grooves $h$ = 0.14 $\mu$m.

\begin{figure}[!h]
\centering
    \includegraphics[scale=0.7]{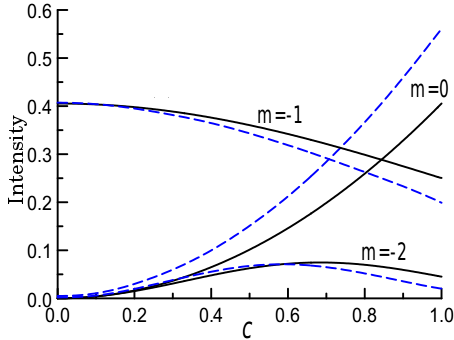}
    \caption{Intensities $I_{m}$ (solid curves, model 1) and $J_{m}$ (dashed curves, model 2).}
\label{fig:intensities}
\end{figure}

Figure~\ref{fig:intensities} shows the intensities of the waves of three diffraction orders $I_{m} =\vert a_{m}\vert ^{2}$, where $a_{m}$ is defined in~\eqref{eq:ampl} (model 1), and $J_{m} = \vert \Psi_{m} (h)\vert ^{2}$ is defined within the dynamical theory (model 2) in relation to the parameter $c = 2hV / dV_{0}$. It can be clearly seen that the intensity of the $0^{th}$ order in model 2 rises with increasing $c$ (i.e. grating velocity) more rapidly than in model 1. The intensity of minus $1^{st}$ order in model 2 decreases faster than in model 1, which is in better agreement with experiments. We used the following data \cite{JETPLett2007}: $V_{0}$ = 4.52 m/s, $h$ = 0.14 mm, d = 5 $\mu$m, $R$ = 6 cm. At maximum rotation frequency of the disk $f_{d}$=105 Hz the velocity $V$ = 39.6 m/s and $c_{max}$ = 0.49.

\begin{figure}[h!]
\centering
    \includegraphics[scale=0.28]{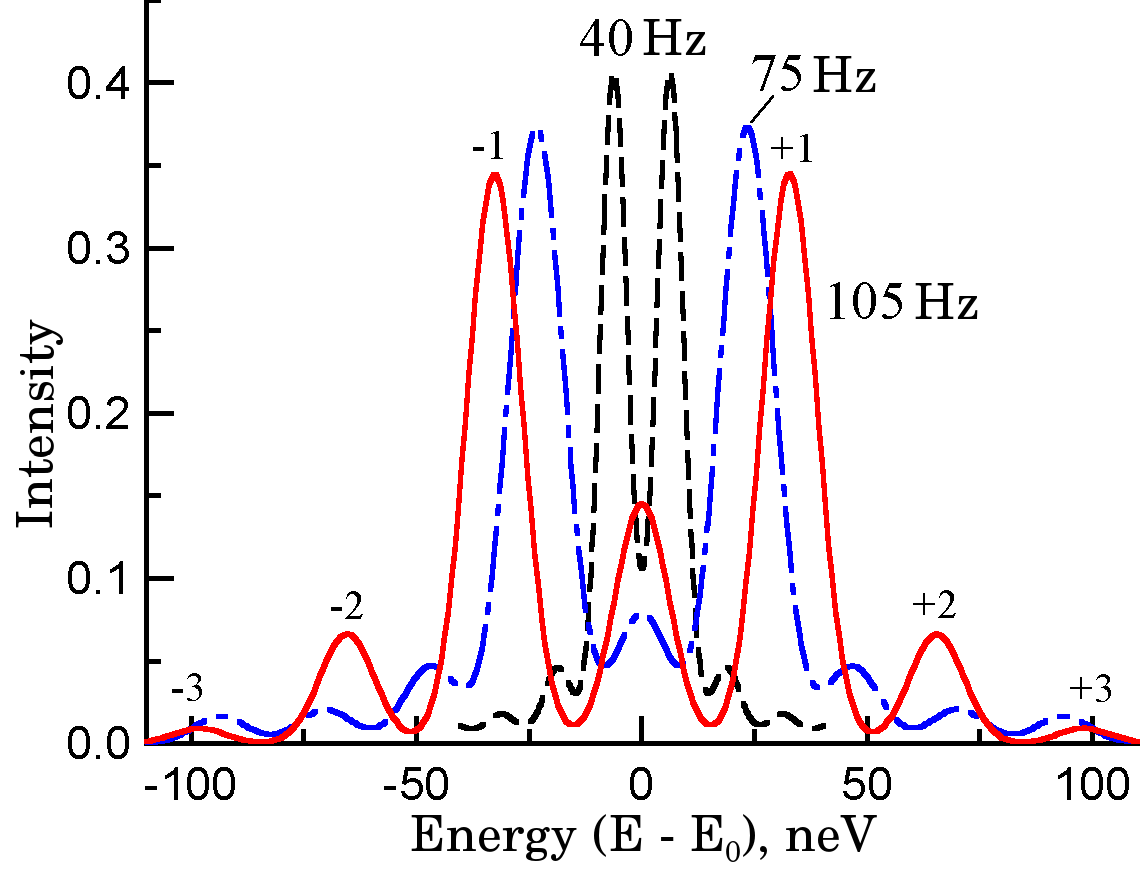}
    \caption{Energy spectra after diffraction at grating rotation frequencies $f_{d}$=40, 75 and 105 Hz.}
\label{fig:espectra}
\end{figure}

Figure~\ref{fig:espectra} shows the energy spectra 

\begin{equation}
I(E)=\frac{1}{\sigma \sqrt{2\pi}}\sum_{m} \vert \Psi_{m}(h)\vert ^{2}exp[-\frac{(E-E_{m})^{2}}{2\sigma ^{2}}],
\label{eq:espectra}
\end{equation}

calculated at different rotation frequencies $f_{d}$, where the dispersion $\sigma^{2} = \sigma_{m}^{2} + \sigma_{a}^{2}$ is determined by the convolution of the spectral functions of monochromator and analyzer of the gravitational spectrometer \cite{PhysLettA2003}, in which the scanning of energy $E = mg\Delta H$ is performed by vertical movement of the analyzer at a distance $\Delta H$, where $g$ is free fall acceleration. The spectral width of the vertical velocities $\Delta V / V_{0}\approx $0.02 \cite{PhysLettA2003,JinrCommun2004,JETPLett2005}.

\begin{figure}[h!]
\centering
      \includegraphics[scale=0.7]{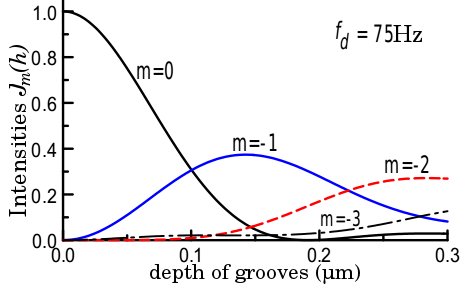}
      \caption{Dependence of $J_{m}$ on the depth of grooves $h$.}
\label{fig:depthdep}
\end{figure}

Above we assumed the depth of the grooves to be fixed ($h$ = 0.14 $\mu$m). It is interesting to see how the intensities of the diffraction orders $J_{m}$ depend on $h$. As is shown on Fig.~\ref{fig:depthdep}, for example, at a depth of h = 0.19 $\mu$m the zero-order intensity is fully suppressed, and the second-order intensity increases. At $h$ = 0.23 $\mu$m the intensities of minus $1^{st}$ and minus $2^{nd}$ orders are equal, and with further increasing $h$ the minus $2^{nd}$ order becomes dominant. Let us note that $J_{m} (h)$ also depends on the frequency $f_{d}$.

Considering UCN diffraction on a grating, we assumed that a plane (i.e. non-divergent) wave falls on it. In fact, in typical experimental conditions \cite{PhysLettA2003,JinrCommun2004,JETPLett2005,JETPLett2007} there is a fairly broad (compared to the vertical velocity $V_{0}$) distribution of the neutron horizontal velocities $v_{x}$. Its absolute value can be as high as 4.5 m/s, which is determined by the value of the boundary velocity for the reflections from the walls of the vertical neutron guide \cite{JinrCommun2004}. To take into account how this distribution affects on diffraction, it is necessary to replace in the above arguments $k_{V}$ with $k_{V} + q_{v}$, where $q_{v} = mv_{x}/\hbar $ - random increments of the x-component of the wave vector, and convolve $\delta$-shaped angular spectrum $F(q)$ in~\eqref{eq:psimovframeint} with the function of the velocity distribution. A specific feature of the diffraction on a moving grating is a decrease of angular divergence $\Delta\theta$ of the incident beam in the grating coordinate system with an increase of the speed of its movement. This improves the condition $\Delta\theta\leq\lambda/d$ for more reliable observation of diffraction. The distribution of horizontal velocities $v_{x}$ with dispersion $\sigma_{v}$ leads to an additional energy broadening $\sigma_{Em}=\hbar\vert q_{m}\vert\sigma_{v}$ of diffraction orders. This broadening is independent on the grating velocity and grows with increasing order number. The typical values of $\sigma_{E1}\sim$1-2 neV.

\bibliography{dindifraction}

\begin{thebibliography}{1}

\bibitem{PhysLettA1994}
Frank A.I. and Nosov V.G.
\newblock {\em Phys.Lett. A}, 188:120, 1994.

\bibitem{PhysLettA2003}
Frank A.I., Balashov S.N., Bondarenko I.V., et~al.
\newblock {\em Phys.Lett. A}, 311:6, 2003.

\bibitem{JinrCommun2004}
Frank A.I., Geltenbort P., Kulin G.V., et~al.
\newblock {\em JINR Communication}, P3-2004-207, 2004.

\bibitem{JETPLett2005}
Frank A.I., Geltenbort P., Kulin G.V., et~al.
\newblock {\em JETP Lett.}, 81:541, 2005.

\bibitem{JETPLett2007}
Frank A.I., Geltenbort P., Jentschel M., et~al.
\newblock {\em JETP Lett.}, 86:225, 2007.

\bibitem{JETPLett2003}
Frank A.I., Geltenbort P., Kulin G.V., et~al.
\newblock {\em JETP Lett.}, 78:224, 2003.

\bibitem{NIMA2009}
Frank A.I., Geltenbort P., Jentschel M., et~al.
\newblock {\em NIM A}, 611:314, 2009.

\end{thebibliography}
\end{document}